\documentclass[%
 reprint,
superscriptaddress,
showpacs,preprintnumbers,
nofootinbib,
 amsmath,amssymb, 
 aps,
 prd,
 bibliography
]{revtex4-1}

\usepackage{cancel}
\usepackage{accents} 
\usepackage{mciteplus,slashed}
\usepackage{amssymb,cancel,amsmath}
\usepackage{dcolumn}% Align table columns on decimal point
\usepackage{bm}% bold math 
\usepackage[caption=false]{subfig}
\usepackage{appendix}
\usepackage{physics}
\usepackage{enumerate}
\usepackage[T1]{fontenc}	
\usepackage{csvsimple}
\usepackage{hyperref}
\usepackage[section]{placeins}
\usepackage[capitalise]{cleveref}
\usepackage{booktabs}
\usepackage{graphicx}
\usepackage{mathrsfs}
\graphicspath{{Figures/}}
\usepackage[utf8]{inputenc}
\usepackage[dvipsnames]{xcolor}
\usepackage[normalem]{ulem}
\newcommand{\be}{\begin{equation}}
\newcommand{\ee}{\end{equation}}
\newcommand{\ba}{\begin{align*}}
\newcommand{\ea}{\end{align*}}
\newcommand{\bpm}{\begin{pmatrix}}
\newcommand{\epm}{\end{pmatrix}}
\newcommand{\bea}{\begin{eqnarray}}
\newcommand{\eea}{\end{eqnarray}}

\newcommand{\benum}{\begin{enumerate}}
\newcommand{\eenum}{\end{enumerate}}
\newcommand{\bi}{\begin{itemize}}
\newcommand{\ei}{\end{itemize}}

\newcommand{\gsim}{\lower.7ex\hbox{$\;\stackrel{\textstyle>}{\sim}\;$}}
\newcommand{\lsim}{\lower.7ex\hbox{$\;\stackrel{\textstyle<}{\sim}\;$}}

\begin{document}

\preprint{\hfill FERMILAB-PUB-20-280-AE-T}

\title{An Active-to-Sterile Neutrino Transition Dipole Moment and the XENON1T Excess}

\author{Ian M. Shoemaker}
\email{shoemaker@vt.edu}
\affiliation{Center for Neutrino Physics, Department of Physics, Virginia Tech University, Blacksburg, VA 24601, USA}

\author{Yu-Dai Tsai}
\email{ytsai@fnal.gov}
\affiliation{Theory Department, Fermi National Accelerator Laboratory, P. O. Box 500, Batavia, IL 60510, USA}
\affiliation{Cosmic Physics Center, Fermi National Accelerator Laboratory, Batavia, IL 60510, USA}
\affiliation{Kavli Institute for Cosmological Physics, University of Chicago, Chicago, IL 60637, USA}

\author{Jason Wyenberg}
\email{jason.wyenberg@coyotes.usd.edu}
\affiliation{Department of Physics, University of South Dakota, Vermillion, SD 57069, USA}

\date{\today}

\begin{abstract}

In this short letter, we find that a magnetic transition dipole moment between tau and sterile neutrinos can account for the XENON1T excess events. Unlike the ordinary neutrino dipole moment, the introduction of the new sterile mass scale allows for astrophysical bounds to be suppressed. Interestingly, the best-fit regions that are compatible with the SN1987A imply either boron-8 or CNO neutrinos as the source flux. We find that sterile neutrinos of either $\sim$ 260 keV or in the $\sim$(500 -- 800) keV mass range are capable of evading astrophysical constraints while being able to successfully explain the XENON1T event rate. The sterile neutrino in the best fit parameter space may have significant effects on big bang nucleosynthesis (BBN). We show the region in which a low reheating temperature of the Universe may allow the BBN constraints to be alleviated.

\end{abstract}

\maketitle 
\section{Introduction \label{sec:1-intro}}

The nature of particle physics beyond the Standard Model (SM) remains unknown. However,
we have two key hints about the nature of new physics: it must account for the non-luminous dark matter (DM), and it must account for neutrino masses. Interestingly, DM direct-detection experiments are sufficiently sensitive to be leading players in searching for novel neutrino interactions that may potentially help solve the mystery of neutrino masses. 

This context makes the recent excess of electron recoil events at XENON1T~\cite{Aprile:2020tmw} all the more intriguing. 
Neutrino magnetic moments were originally studied by the XENON1T collaboration as potential explanations to the excess (axions, dark photons, and other DM proposals were also discussed in ~\cite{Takahashi:2020bpq,Kannike:2020agf,Alonso-Alvarez:2020cdv,Amaral:2020tga,Fornal:2020npv,Boehm:2020ltd,Harigaya:2020ckz,Bally:2020yid,Su:2020zny,Du:2020ybt,DiLuzio:2020jjp,Bell:2020bes,Chen:2020gcl,Dey:2020sai,Choi:2020udy,AristizabalSierra:2020edu,Buch:2020mrg,Paz:2020pbc,Lee:2020wmh,Cao:2020bwd,Robinson:2020gfu,Khan:2020vaf,Primulando:2020rdk,Nakayama:2020ikz,Jho:2020sku,Bramante:2020zos,Baryakhtar:2020rwy,An:2020bxd,Zu:2020idx,Gao:2020wer,Budnik:2020nwz,Lindner:2020kko,Bloch:2020uzh,DeRocco:2020xdt,Dent:2020jhf,Zioutas:2020cul,McKeen:2020vpf,Coloma:2020voz,An:2020tcg,DelleRose:2020pbh,Ge:2020jfn,Bhattacherjee:2020qmv,Dessert:2020vxy,Chao:2020yro,Cacciapaglia:2020kbf,Ko:2020gdg,Gao:2020wfr,Alhazmi:2020fju,Baek:2020owl,Li:2020naa,Chigusa:2020bgq,Miranda:2020kwy,Benakli:2020vng,Okada:2020evk}). However, the couplings found tend to exceed the bounds from various astrophysical systems. In this paper, we highlight a neutrino dipole-portal interaction that can account for the signal, while evading astrophysical bounds (though still could be subject to cosmological bounds). This results from the introduction of a new mass scale to the neutrino interaction. 
%checked

%%%%%%%%%%%%%
\begin{figure}[b!]
\includegraphics[angle=0,width=.45\textwidth]{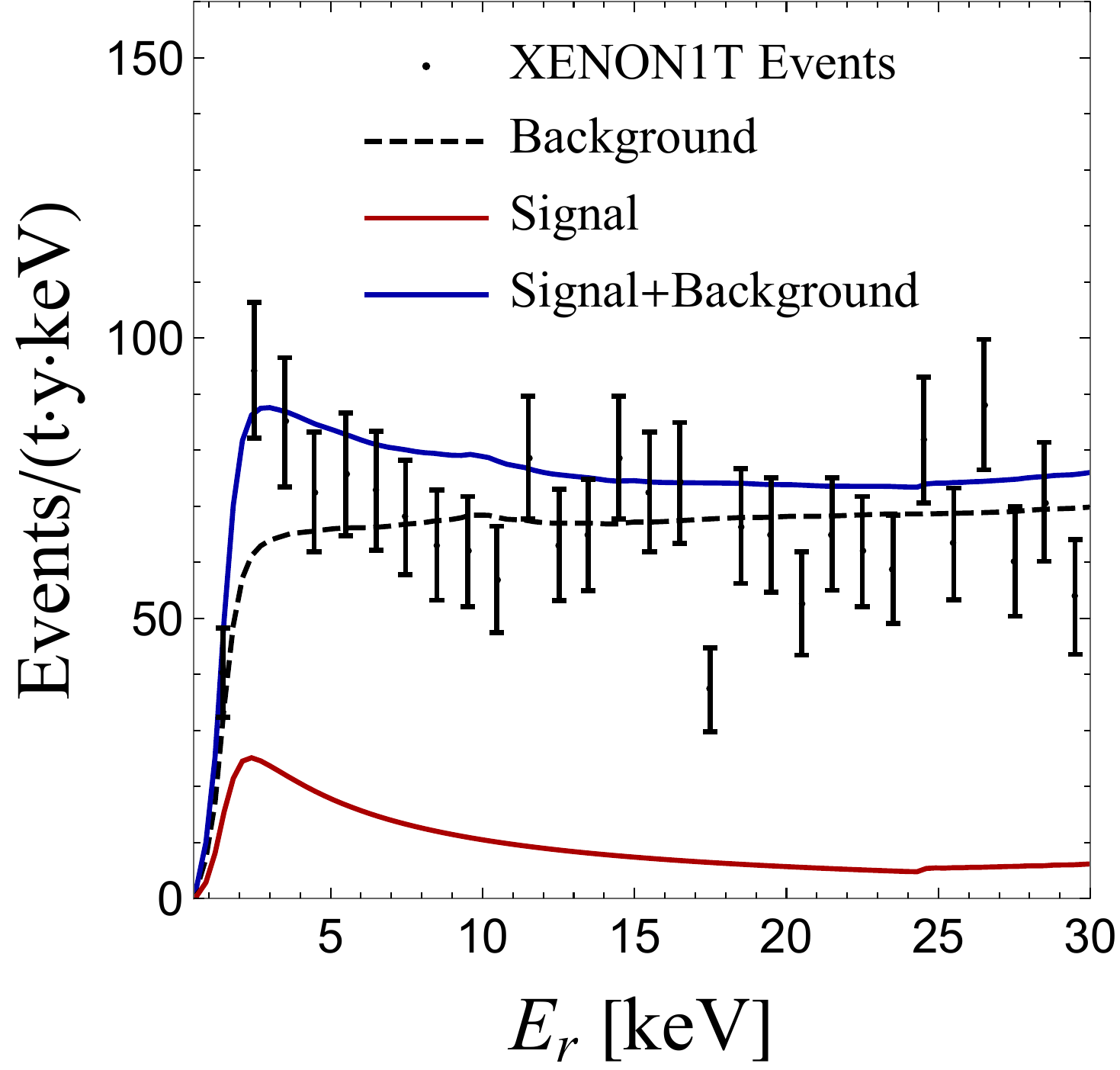}
\caption{Dipole portal best fit signal spectrum at XENON1T with $m_{4} = 640$ keV and $d = 2.2 \times 10^{-9}~\mu_{B}$. The background is shown in dashed black, the signal is solid red, and the signal plus background is shown in solid blue. Included in these event rates are the energy-dependent signal efficiency.}
\label{fig:fluxes}
\end{figure}
%%%%%%%%%%%%%

The most commonly studied models accounting for neutrino masses introduce right-handed sterile neutrinos, $N$, via the interaction $\mathcal{L} \supset HNL$. However, it is important to stress that such singlet states need not dominantly interact with the SM through this particular operator. Viable scenarios exist in which the dominant interaction comes from an active-to-sterile dipole moment, sometimes referred to as the ``neutrino dipole portal,''
\be 
\mathcal{L}  \supset d \left(\bar{\nu}_{L}  \sigma_{\mu \nu} F^{\mu \nu} N \right) + h.c.,
\label{eq:Lagr}
\ee
where $F_{\mu \nu}$ is the electromagnetic field strength, $\sigma_{\rho\sigma}=\frac{i}{2}[\gamma_\rho,\gamma_\sigma]$, $\nu_{L}$ is the SM neutrino, and the coefficient $d$ with units of $(\rm{mass})^{-1}$ controls the strength of the interaction. This transition dipole moment has been studied in the context of MiniBooNE~\cite{Gninenko:2009ks,Gninenko:2010pr,McKeen:2010rx,Masip:2011qb,Gninenko:2012rw,Masip:2012ke,Magill:2018jla,Bertuzzo:2018itn}, and future projected bounds have been studied for IceCube~\cite{Coloma:2017ppo}, SHiP~\cite{Magill:2018jla}, and direct-detection experiments~\cite{Shoemaker:2018vii}.

%%%%%%%%%%%%%
\begin{figure}[t!]
\includegraphics[angle=0,width=.49\textwidth]{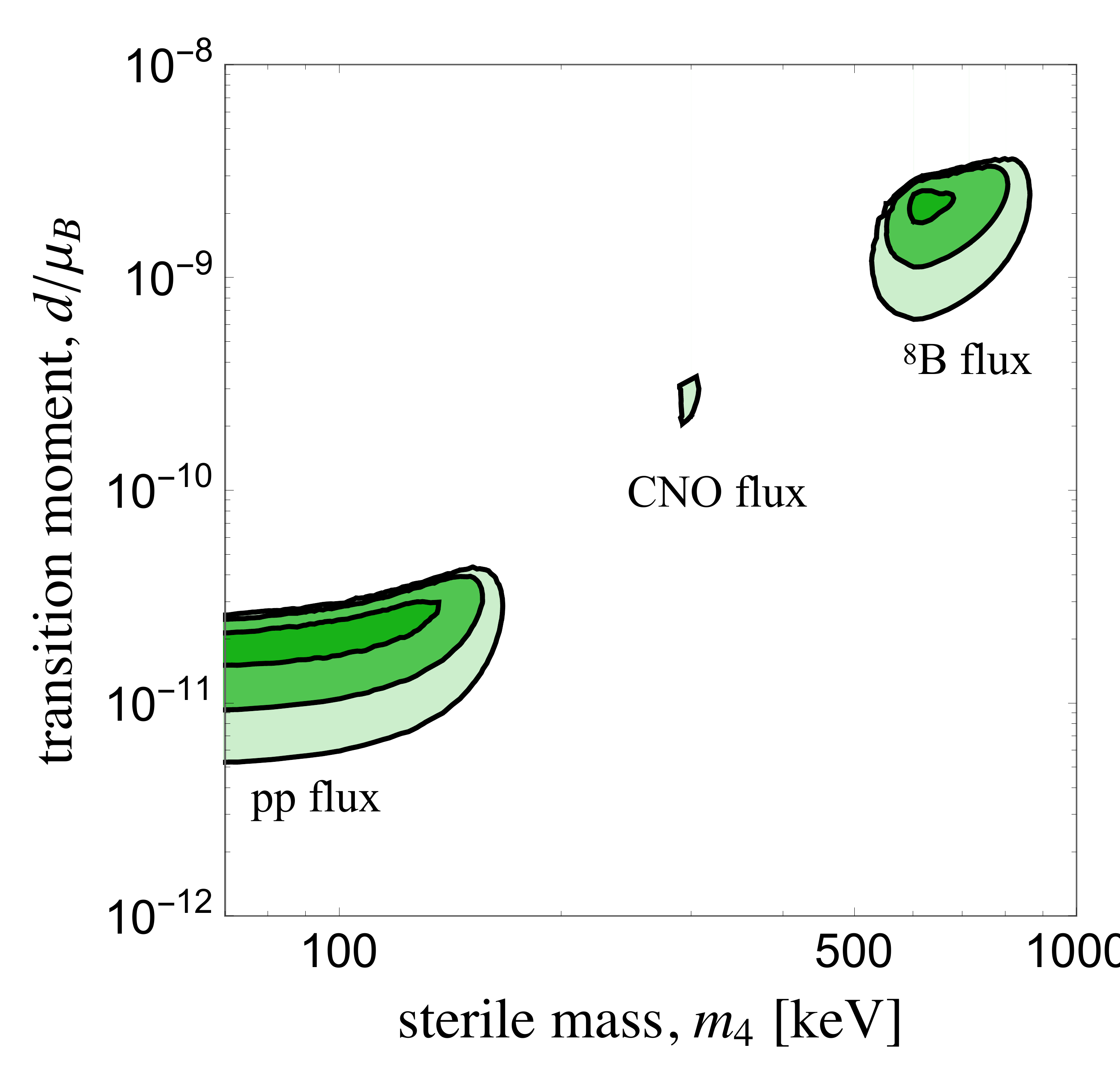}
\caption{We display the best-fit contours (1,2,3)$\sigma$ in the dipole coupling-mass plane. Each region originates from the labeled solar neutrino fluxes.} 
\label{fig:fit}
\end{figure}
%%%%%%%%%%%%%

Note that this operator can be induced through loop processes with the ordinary HNL operator, and in most cases, mixing between $N$ and SM neutrinos would also be induced. 
We focus our attention on the operator in Eq.~(\ref{eq:Lagr}) as a simplified consideration. We also introduce a Dirac mass for $N$, denoted $m_{4}$, and avoid the Majorana mass-term for complications discussed in \cite{Magill:2018jla}.
In addition, a small mixing angle would not affect our result, and the constraints are dependent on the size of the mixing. 
The tau mixing is not strongly constrained for a $\sim$ 100 keV sterile neutrino ~\cite{Helo:2011yg,Bryman:2019bjg,Dib:2019tuj,Kim:2019xqj,Bryman:2019ssi}. We leave a more complete consideration including the mixing and the potential complications of introducing a Majorana mass term for a future work \cite{precite}.

We will show that up-scattering from $\nu$ to a heavy sterile neutrino $N$ (with a Dirac mass $m_{4}$) within the XENON1T detector can plausibly explain the excess while simultaneously evading astrophysical and other terrestrial bounds. We find that $\rm pp$, CNO, and $^{8}\rm B$ solar neutrino fluxes lead to three separate best-fit regions. While the pp region is known from the XENON1T analysis~\cite{Aprile:2020tmw}, we show that this extends to masses of 100 keV. 
Furthermore, CNO and $^{8}\rm B$ neutrino fluxes respectively allow novel $\sim$ 260 keV and $\sim$ 600 keV sterile mass solutions assuming this minimal dipole operator. We note that the first direct experimental evidence for neutrinos from the CNO cycle was just reported by Borexino~\cite{Agostini:2020lci}. 

The remainder of this paper is organized as follows. In Sec.~\ref{sec:calc} we introduce the main framework for computing event rates at XENON1T. In Sec.~\ref{sec:results} we discuss the results of our fit to the XENON1T data under a neutrino dipole interpretation. In Sec.~\ref{sec:constraints} we review existing constraints on the model focusing on direct detection, stellar energy loss, supernovae, and BBN. Finally we conclude the paper in Sec.~\ref{sec:concl} and discuss future probes of the model.
\newline
%to here!

\section{Energy Deposition via Neutrino Dipole Portal}
\label{sec:calc}

As studied in~\cite{Shoemaker:2018vii}, incoming solar neutrinos can upscatter to the heavy sterile state $N$ in the detector volume of a direct-detection experiment. They will inevitably decay as well through $N\rightarrow \nu + \gamma$, but in the cases of interest for the present, the decay length is much longer than the detector dimensions.

To estimate the event rate, we consider the up-scattering cross section:
\begin{widetext}
\bea
\label{eq:magneticmomentcrosssection}
\frac{d\sigma_{\nu e \rightarrow Ne}}{dE_R} &=& d^2\alpha  \Bigg[\frac{1}{E_R}-\frac{m_{4}^2}{2E_{\nu}E_{R}m_e}\Bigg(1-\frac{E_{R}}{2E_\nu}+\frac{m_e}{2E_\nu}\Bigg)-\frac{1}{E_\nu}+\frac{m_{4}^4(E_{R}-m_e)}{8E_\nu^2E_{R}^2m_e^2}\Bigg].
\eea
\end{widetext}
Here, $d$ is the coupling constant defined in Eq. \ref{eq:Lagr}, $\alpha$ is the fine structure constant, $m_e$ is the electron mass, $m_{4}$ is the mass of the heavy sterile neutrino,  $E_\nu$ is the incoming neutrino energy, and $E_{R}$ is the electron-recoil energy.

The electron-recoil spectrum of the up-scattering events can be determined as
\be
\label{eq:magneticmomentrate}
\frac{dR^i}{dE_R} = {\rm MT} \times\int_{E_{\nu}^{min}}\frac{d\Phi_{\nu}^{i}}{dE_{\nu}}\frac{d\sigma_{\nu e\rightarrow Ne}^i }{dE_R}(E_{\nu},E_{R}\big)dE_{\nu},
\ee
where $i$=$\tau$ for the tau neutrino flavor only (unlike the case in \cite{Shoemaker:2018vii}).
In addition, $\Phi_\nu$ is the solar neutrino flux and $\rm MT$ is the exposure.
The minimum energy of the incoming neutrino to up-scatter to the $m_{4}$-mass state, yielding electron recoil with an energy $E_{R}$, is
\be
\label{eq:minimumenergy}
E_\nu^{{\rm min}}(E_{R})=\frac{m_{4}^2+2m_eE_R}{2\Big[\sqrt[]{E_R(E_R+2m_e)}-E_R \Big]}.
\ee

We also include the energy-dependent efficiency given in Fig. 2 of Ref.~\cite{Aprile:2020tmw}, and include their background estimations in our fits.

%%%
\section{Results}
\label{sec:results}
%%%%%%%%%%%%% 

We display our main results in Fig.~\ref{fig:fit} assuming only a $\nu_{\tau}$ coupling for the dipole interaction in Eq.~\ref{eq:Lagr}. We follow the method of \cite{Billard:2013qya} and utilize a Log-Likelihood Profile method to quantify a discovery significance. The likelihood of the XENON1T results being produced by background only is compared to the likelihood from the background plus hypothesized signal, and the resulting test statistic is used to generate a discovery significance for each point in the $m_{4}$, $d$ parameter space. Because of the additional degree of freedom provided by the new mass scale, the asymptotic form of the test statistic follows a $\chi^2$ distribution with two degrees of freedom as described in \cite{cowan_asymptotic_2011}. The discovery significance may be lower than we have stated because of the look-elsewhere effect (due to our selective sampling of data in the XENON1T low energy bins only), but we leave a more detailed statistical analysis for further research. Along with the XENON1T contours we display the SN1987A~\cite{Magill:2018jla} and previous XENON1T bounds based on nuclear recoil data~\cite{Shoemaker:2018vii}. At larger couplings, the dipole is constrained by LEP~\cite{Magill:2018jla} and DONUT~\cite{Schwienhorst:2001sj,Coloma:2017ppo}. 

The equivalent plot for $\mu$-flavor coupling includes strong bounds from CHARM-II at the $3\times 10^{-9}~\mu_{B}$ level. Moreover, Borexino probes up to $m_{4}\simeq 230$ keV~\cite{Coloma:2017ppo} at the $2.8 \times 10^{-11}~\mu_{B}$ level for all flavors~\cite{Borexino:2017fbd}. Thus the CNO and boron-8 region will survive in the muon-coupling case as well. 

Next, let us understand the origin of the separate islands in Fig.~\ref{fig:fit}. Given that the ordinary neutrino magnetic moment can explain the XENON1T event rate via pp neutrinos, we can normalize our expectation for the dipole portal coupling via CNO as:
\be
\Phi_{pp}~ \mu_{\nu}^{2} =\Phi_{N,O} ~ d^{2}
\ee
where $\mu_{\nu} \simeq 2 \times 10^{-11}~\mu_{B}$ is the active neutrino magnetic moment needed for the XENON1T excess, while $d$ is the new active-sterile magnetic moment (see Eq.~\ref{eq:Lagr}), and $\Phi_{pp}$ and $\Phi_{N,O}$ are the pp and N,O solar neutrino fluxes. Thus given that $\Phi_{pp} /\Phi_{N,O} \simeq 10^{2}$, we expect $d\simeq \mu_{\nu}~ \sqrt{\Phi_{pp}/\Phi_{N,O}}  \simeq 10^{-10}~\mu_{B}$.  

Further, the maximum sterile mass probed can be found from
\be 
m_{4,max}^{2} = 2\left[E_{\nu}\sqrt{E_{R}(E_{R}+2m_{e})}-E_{R}(E_{\nu}+m_{e})\right].
\ee
For the MeV-scale CNO neutrinos, the largest mass probed for the 2-3 keV recoil energies of interest is $m_{4,max}^{\rm CNO} \simeq 290$ keV, in good agreement with the fit in Fig.~\ref{fig:fit}. The same logic can be applied to the $^{8}\rm B$ flux, from which we estimate $m_{4,max}^{^{8}\rm B} \simeq 920$ keV and $d\simeq 3 \times 10^{-9}\mu_{B}$.

\section{Constraints}
\label{sec:constraints}
%%%%%%%%%%%%%
\begin{figure}[t!]
\includegraphics[angle=0,width=.5\textwidth]{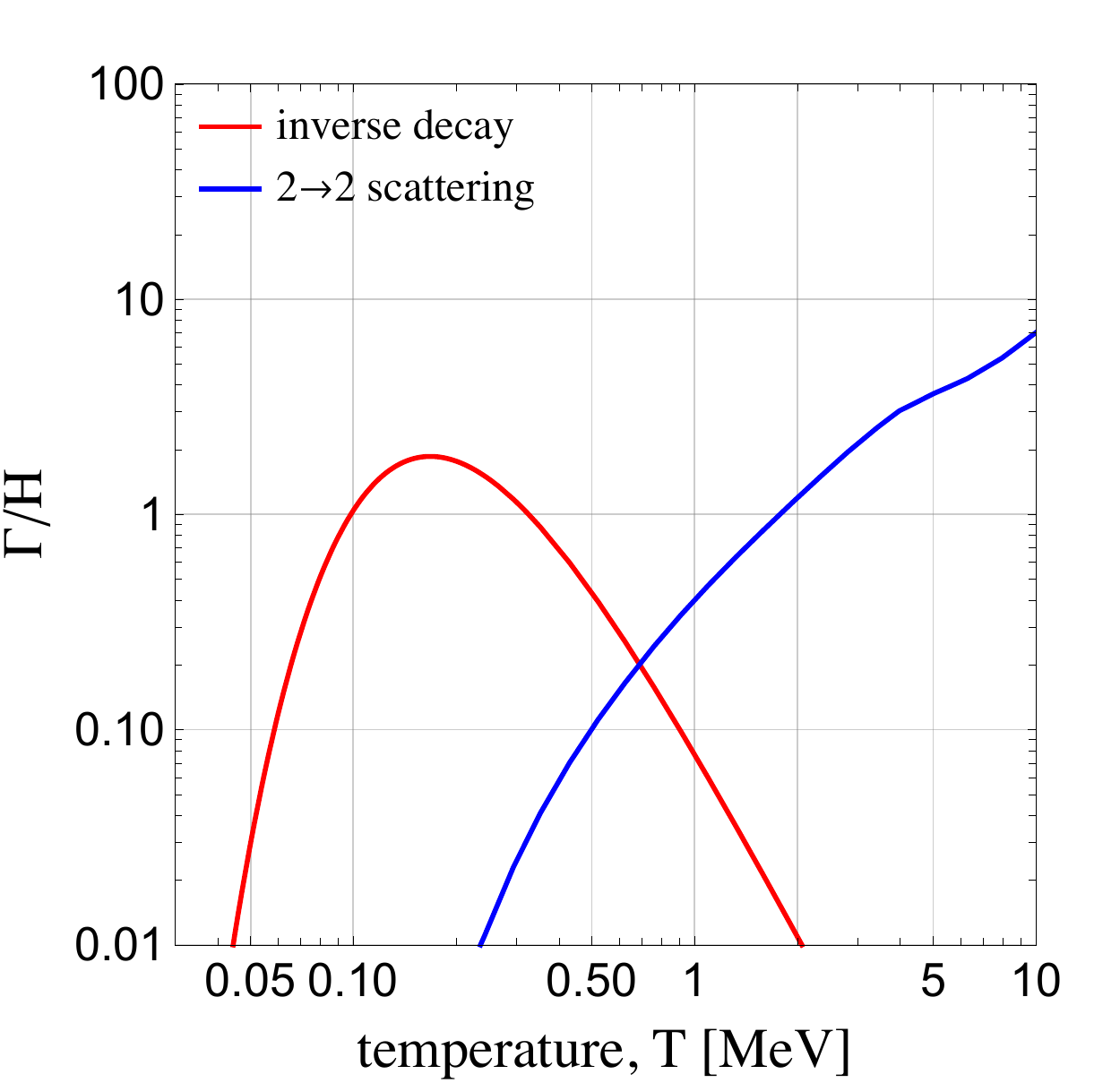}
\caption{Comparison of the thermalization rates (inverse decay and 2$\rightarrow$ 2 upscattering/synthesis (see text for descriptions), based on the thermal-averaged cross-sections and number density to the temperature-dependent Hubble rate, for the near best-fit benchmark parameter point from Fig.~\ref{fig:fit}, $m_{4} = 600$ keV and $d=10^{-9}~\mu_{B}$.}
\label{fig:BBN}
\end{figure}
%%%%%%%%%%%%%

\subsection{Xenon 1T NR constraint}

Previous work utilized XENON1T's nuclear recoil search for DM~\cite{Aprile:2018dbl} to place conservative constraints on the neutrino dipole portal interaction~\cite{Shoemaker:2018vii}. In that case, the low-energy solar neutrinos scatter coherently such that there is a $\sim Z^{2}$ enhancement in the cross-section. This analysis~\cite{Shoemaker:2018vii} specifically vetoed events if the produced heavy sterile decayed inside the detector volume since this would produce an altered ionization/scintillation signal. Despite this reduction in event rate, the bounds derived from the XENON1T nuclear recoil data~\cite{Aprile:2018dbl} were stronger than existing bounds on the tau-flavored dipole portal coupling for sterile masses less than 10 MeV.

\subsection{Stellar Energy Loss}

The heavy sterile neutrino $N$ we consider can carry energy when thermally produced in the stellar systems, affecting the energy loss, thermal conductivity, and eventually time evolution of well known stellar populations (see, e.g.,  \cite{Raffelt:1994ry,Arceo-Diaz:2015pva,daz2019constraint}). 
When $m_{4}>T_{\rm star}$ ($T_{\rm star}$ is the temperature of the star) the thermal-averaged energy loss is proportional to $\exp(-m_{4}/T_{\rm star})$ \cite{Pospelov:2017kep}.
Taking the powerful production of $N$ in the red-giants into account, our parameter space of interest of a heavy sterile neutrino $N$ below $\sim$ 250 keV would be constrained, as indicated by the red dashed vertical line in Fig. \ref{fig:cons}.

\subsection{Supernova 1987A}

The constraints from Supernova 1987A (SN1987A) on a dipole portal heavy sterile neutrino is conducted in \cite{Magill:2018jla}. 
The constraint in our parameter space of interest is enclosed by an upper limit and a lower limit.
The lower limit is set by the minimal emission of sterile neutrino $N$ that would carry out enough energy to affect the standard supernovae cooling through neutrinos from the core. The relevant production processes include neutrino upscattering to $N$ with electrons/positrons/protons, $e^+ e^-$ annihilating into $N \bar{\nu}$, and neutrino-photon inverse decay.
The higher limit of the bound is set by the "trapping" of the sterile neutrino $N$, meaning that the energy carried by $N$ can be recycled and re-emitted within the “neutrinosphere” (an isotherm-sphere within which the neutrino is diffusive rather than free-streaming) \cite{Chang:2016ntp}, and the constraint from supernova cooling can be avoided with large enough coupling. The relevant processes for this "trapping" consideration are $N$ downscattering to neutrinos with electrons/positrons/protons, $N-\bar{\nu}$ annihilation to electron pairs, $N$ decay, and gravitational trapping (which kick in at above $\sim$ 300 MeV). The constrained regime is shown in Fig. \ref{fig:cons}.

Notice that the SN1987A bound is subject to large uncertainties in both the supernovae property, the SN1987A measurement, and also the cooling model \cite{Bar:2019ifz}. 
Improvements of the SN1987A bound consideration is possible, both for sub-MeV and above-MeV regimes. One thing that can be taken into account is the complicated flavor composition and energy spectra of neutrinos during supernovae processes (see, e.g., \cite{Fischer_2010}). Given that our sterile neutrino is only tau coupled, such consideration may affect the supernovae bound. Further analysis can be performed with a more detailed study \cite{precite}.

\subsection{BBN}

Dark-sector particles of mass around or below a MeV that couple to the SM sector could be subject to strong big bang nucleosynthesis (BBN) constraints {\cite{Berezhiani:2012ru,Boehm:2013jpa,Krnjaic:2019dzc}}. Previous work~\cite{Magill:2018jla} found that BBN constraints on the heavy sterile neutrino with a dipole portal are sensitive to the reheating temperature for a broad range of sterile masses. If $N$ is thermally populated in the early universe, it could have various effects on BBN. In particular, it could contribute to $N_{\rm eff}$ as a semi-relativistic particle during BBN, and it could decay to photons and neutrinos, affecting their temperatures (also affecting $N_{\rm eff}$) as well as light-element production \cite{Berezhiani:2012ru,Boehm:2013jpa}.

The major thermalization processes of $N$ are the
 $2 \rightarrow 2$ processes $e^{+} + e^{-} \rightarrow N+ \bar{\nu}$ (synthesis) and  $e^{-} + \nu \rightarrow e^{-} + N$ (up-scattering), and neutrino-photon inverse decay, $\nu+\gamma \rightarrow N$. Since the 2 to 2 processes scale as $\Gamma_{ 2\rightarrow 2}\;\undertilde{\propto}\;d^{2} T^{3}$, $N$ will inevitably be thermalized when the SM temperature $T$ is much higher than the energy scale $d^{-1}$. Under the assumption of high reheating temperature ($T_{\rm RH}$ much larger than $d^{-1}$), the sub-MeV region of interest would be strongly constrained.

\begin{figure}[t!]
\includegraphics[angle=0,width=.5\textwidth]{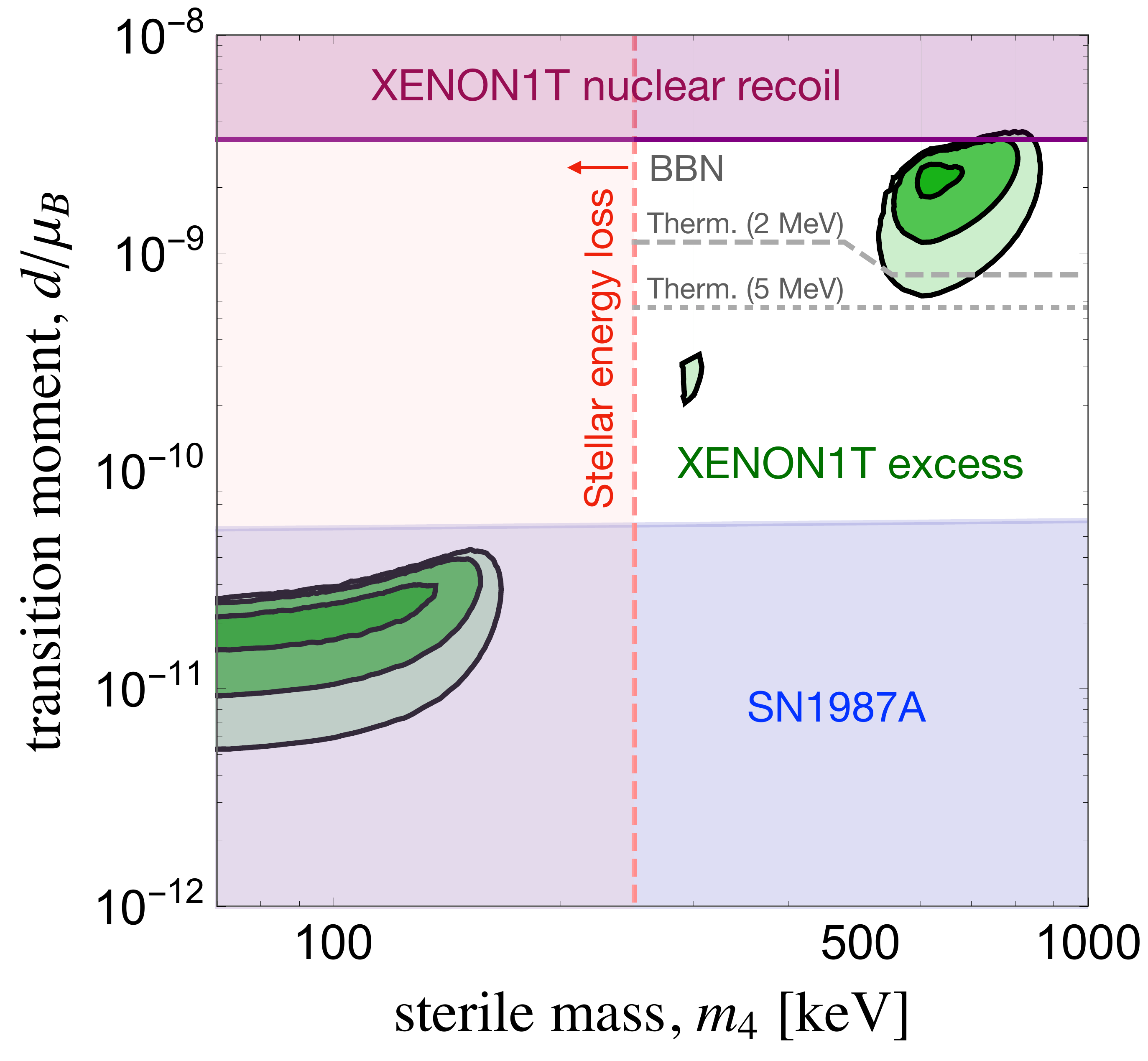}
\caption{
Along with the XENON1T best-fit contours, we display the SN1987A~\cite{Magill:2018jla}, stellar energy loss, and XENON1T nuclear recoil bounds~\cite{Shoemaker:2018vii}. 
We also use dashed/dotted gray curves to indicate the parameter space in which thermalization (thermal processes exceeding the Hubble rate) could happen at a temperature below 2 MeV (region above the upper dashed-gray) and 5 MeV (region above the lower dotted-gray), see texts for detail.
}
\label{fig:cons}
\end{figure}
%%%%%%%%%%%%%

However, one can consider cosmological scenarios with a very low reheating temperature \cite{Moroi:1999zb,Hannestad:2004px,Kohri:2005wn,Pradler:2006hh,deSalas:2015glj,Berlin:2018bsc}, around 2 to 5 MeV, and determine if $N$ will be thermalized below such temperatures. In the parameter space that $N$ is not thermalized, the strong BBN bound can be alleviated.
In Fig.~\ref{fig:BBN}, we show the ratio of the major thermalization processes to the Hubble rate, for a benchmark point in the best fit region for $^{8}\rm B$.
These rates are calculated through proper thermal-averaging (briefly described in \cite{Magill:2018jla}). We also checked that the numerically calculated thermal-averaged cross-section and the analytical cross-sections taking into account boost-factors from the temperature match quantitatively.
As one can see, the neutrino-photon inverse decay can play a major role in thermalization with a low-reheating temperature, as it has a resonance region at around $T \sim m_{4}/2.$ In Fig.~\ref{fig:BBN}, one can see the behavior of $\Gamma/H$ in different SM temperatures.
Note that $\Gamma/H\sim 1$ is just an estimation of thermalization. Detailed consideration should be conducted to determine the actual thermalization conditions.

In Fig. \ref{fig:cons}, we display the calculated thermalization curves for the parameter space of interest.
Above the upper (lower) dashed gray curves is the parameter space in which the $N$ thermalization rate exceeds the Hubble rate below 2 MeV (5 MeV) temperatures.
The BBN constraints may be alleviated below these curves.
Note that the $N$ thermalization processes mentioned above also provide new channels for electron/photon and neutrino sectors to thermalize (even if the $N$ itself is not being fully thermally populated).
However, the parameter space below the "thermalization curves" in Fig. \ref{fig:cons} still indicate that these interaction rates never exceed the Hubble rate under the low reheating temperatures we have considered (2 and 5 MeV), and that these rates may not be strong enough to directly affect neutrino decoupling or BBN.
We find some of the best-fit regions for $^{8}\rm B$ flux are covered by 2 MeV or 5 MeV thermalization curves, and the small island of CNO flux-favored region can avoid thermalization.
Given the borderline case for the $^{8}\rm B$ region, we believe that a detailed analysis should be conducted to determine the full effects of this portal on neutrino decoupling and BBN. We leave this consideration for a future work \cite{precite}.

\section{Conclusions}
\label{sec:concl}

We have shown that the neutrino dipole portal can successfully reconcile a solar neutrino origin with astrophysical and terrestrial bounds with the introduction of a new mass scale sterile neutrino. Future probes of the model include: scintillation-only LXe data~\cite{Aprile:2019xxb}, SHiP~\cite{Magill:2018jla}, DUNE, and a detailed BBN analysis. We further note that direct atmospheric production of sterile neutrinos via the dipole operator may lead to additional constraints~\cite{Bueno:2013mfa}. Although such models may induce mass mixing with the active neutrinos, these bounds are weak~\cite{Helo:2011yg,Bryman:2019bjg,Dib:2019tuj,Kim:2019xqj,Bryman:2019ssi} for the tau-flavor couplings we focus on.

\section*{Acknowledgements}

We thank Nikita Blinov, Pilar Coloma, Gordan Krnjaic, Maxim Pospelov, Evan Shockley, and Luca Vecchi for useful discussions.
The work of IMS is supported by the U.S. Department of Energy
under the award number DE-SC0020250.
This document was prepared by Y-DT using the resources of the Fermi National Accelerator Laboratory (Fermilab), a U.S. Department of Energy, Office of Science, HEP User Facility. Fermilab is managed by Fermi Research Alliance, LLC (FRA), acting under Contract No. DE-AC02-07CH11359.

\vspace{5cm}

\bibliography{ms.bib}
\end{document}